\documentstyle[twocolumn,aps]{revtex}
\begin{document}
\draft
\title{
\Large\bf Reconciling Inflation with Openness} 
\author{Luca Amendola$^{1}$, Carlo Baccigalupi$^{1,2}$ and
Franco Occhionero$^{1}$}

\address{$^{1}$
Osservatorio Astronomico di Roma,
Viale del Parco Mellini 84, 00136 Roma, Italy
}
\address{$^{2}$ University of Ferrara,
Via Savonarola 9, 44100 Ferrara, Italy}

\maketitle

\vspace{2.cm} 
\begin{abstract}

It is already understood that
the increasing observational evidence for an open Universe can be
reconciled with inflation if our horizon is
contained inside one single huge bubble nucleated during the inflationary
phase transition. In this frame of ideas, we show here 
that the probability of living in  a bubble
with the  right $\Omega_0$  
(now the observations require $\Omega_0\approx\, .2\,$)
can be comparable with unity, rather than infinitesimally small. 
For this purpose   we modify both quantitatively 
and qualitatively an intuitive  toy model   based
upon fourth order gravity.   As this scheme 
can be implemented in canonical
General Relativity as well (although then the inflation driving
potential must be designed entirely {\it ad hoc}),
inferring from the observations that $\Omega_0<1$ not only does not conflict
with the inflationary paradigm, but rather 
 supports therein the occurrence of a primordial
phase transition.

\end{abstract}
 
\vspace{.3cm}
 

{\bf Introduction.}    An open Universe, 
i.e. without enough matter to halt eventually its expansion, 
($\Omega_0\approx 0.2 $), agrees with most astronomical observations 
(see e.g. Ref. \cite{BLD})
and  with their interpretation. For example, in connection with
the formation of large scale structure in the CDM
scenario, it gives the best fit to the observed clustering (see e.g.
Ref. \cite{CFA2}) yielding also the required \cite{DEFW} power 
on the large scales; 
it explains the dynamics of bound objects on relatively small scales 
(see e.g.
Ref. \cite{PEE}); it also increases the age of the Universe, alleviating 
the conflict  with the age of  globular clusters (see e.g. \cite{BHM}); 
finally, 
it is  in better agreement with direct geometrical estimates from
radio source number counts (see e.g. Ref. \cite{PEE} \cite{RP}).
A low density Universe is now preferrable even for theorists
(e.g. \cite{OS}, \cite{Gaeta}) when they essay to explain
the small scale anisotropies measured by the {\it COBE} satellite
in the Cosmic Microwave Background (CMB). Quite naturally,
as  the  flatness prediction,  $\Omega_0=1$, is a basic paradigm 
of inflation, one may resort to  the addition of a vacuum energy, 
$\Omega_0=\Omega_{\rm CDM}+\Omega_{\Lambda}=1$,
(for an early suggestion see e.g. Ref. \cite{MST})
 and build a theory for $\Lambda$, see e.g. Ref. \cite{Gaeta}.
Nevertheless, that $\Omega_0$ may be less than one
 is certainly a stimulating  challenge of
modern cosmology, notwithstanding the obvious caveat that nothing is certain
because i)  the present observations  
may be too limited to be representative of the whole
Universe (and that higher values of $\Omega_0$ might be found at larger 
scales)
and ii)  the job of interfacing  theory and observations
for  the CMB perturbations  is   still in its infancy
(see e.g. Ref. \cite{Pogosyan} for another solution).
On the other hand,
it is also possible to choose the initial conditions in
inflation so as to give $\Omega_0\approx 0.2$ today, either by starting with
an extremely small density parameter at the beginning of
inflation, or by assuming that inflation lasted less than
60 $e$-foldings or so. Both possibilities, however,
enter in conflict with the very spirit of inflation because they  
introduce the fine-tuning of the initial conditions that
inflation  overcame and we certainly do not want to reintroduce; 
moreover, there would be also a conflict with the 
microwave background isotropy
\cite{KTF}.

In this work,   we move  instead from the fundamental
notion \cite{COLEMAN} \cite{GOTT} that the inside of a 
primordial bubble
nucleated for quantum tunneling from a false vacuum (FV) to a true vacuum 
(TV)
looks like an open Universe in an  external Universe  already 
totally flat.  Recently substantial progress (upon which we build)
has been achieved  along this line  in two different ways.
The first is the single-bubble scenario \cite{GW}
\cite{JAP} \cite{BGT} in one field inflation,  where
the  identical bubbles inflated for about 60 $e$-foldings after nucleation:
our visible Universe is contained inside one of these
 bubbles, and appears to be locally open. 
The second proposal  \cite{LIN} is the many-bubble scenario in
 two-field inflation  where one field drives the
inflationary slow-rolling and the other undergoes a quantum tunneling in a
direction orthogonal to the former, generating  bubble-like open 
Universes, with all possible density parameters, from zero to unity.
 Then, there is no reason to expect  a preferential value of $\Omega_0$: 
it must then be  argued that possibly quantum cosmology will explain why
we live in an $\Omega=0.2$ Universe.

The model of this work implements also a many-bubble scenario,
 exploits fully  the assets of two field inflation \cite{AF}
  and  has two useful features:  the peak of the 
bubble nucleation i) can be placed at  any observed  $\Omega_0<1$, and
ii) can be made narrow enough for our Universe to be regarded as typical.
Furthermore, the absolute probability 
of residing in a bubble  (whatsoever) may be made
comparable to one (at least until some costraint is found), 
so that the use of the anthropic principle may be largely avoided. 
In our model bubble nucleation is made to end abruptly: thereafter,
the external space (embodied in the residual fraction of false vacuum)
undertakes a classical double
inflationary tour exactly like in the literature \cite{DI} that seeks
a break in the canonical featureless perturbation power spectrum. 
In fact, this is an anomalous two field inflation in which 
the one scalar field  present  (the other in reality is disguised as gravity)
is exploited twice, quantistically (for the tunneling)   
and classically (for the second slow rolling).

Our model, already
introduced  \cite{OA} to produce large scale power   in the CDM scenario
out of the remnants of the primordial phase transition (see also \cite{AO}), 
contains now  a result of \cite{BGT} that specifies
how to link the bubble's $\Omega_0$  to the bubble's nucleation epoch $N$
(number of $e$-foldings between nucleation and end of inflation),
  
	\begin{equation} \label{OMEGA0}
\Omega_0(N)=\left[  1+4\exp 2\,(N_H-N) \right]^{-1}\;,
	\end{equation}
where $N_H$ corresponds to the horizon, i.e. is  
fixed by the request that the largest observable scale, $L_H=2c/H_0$,
 crossed out the horizon $N_H$ $e$-foldings before the end
of inflation (and  is close to 60 for standard
cosmological values \cite{KT}). Notice that
according to (\ref{OMEGA0}), for $N\rightarrow \infty\;, \Omega_0
\rightarrow 1$ as expected because the bubble is born with $\Omega_0=0$
and evolves toward flatness with time. Also, notice the other
coincidence that bubbles born at $N_H$ have today $\Omega_0=.2\,$: this is
the main difficulty in model building, because a horizon sized bubble  
 (or even smaller, were the need to arise) can easily be seen.  

Our procedure is the 
following: given any value of $\Omega_0$ we determine through
(\ref{OMEGA0}) the corresponding $N(\Omega_0)$ and we end abruptly
at that $N_E=N$ the bubble production via a feature in the potential 
(see below) that increases manifolds the Euclidean action. 
The Universe contains then only bubbles that have been generated earlier,
$N>N_E$, in higher number for lower $N$ 
and have now attained all the $\Omega_0(N_E)<\Omega_0<1$.
This break is also new with respect to our work of \cite{OA}. 
In  Ref. \cite{LIN}, the lack of completion of the phase transition
is dictated by the need  to prevent us from seeing
our bubble's walls; here, it solves the same problem,
but in a novel way, that may be useful in future improved work:
 a virtuous (or cunning) potential takes care of inflating the bubble size
 beyond the horizon after the bubble interior has become radiation
 dominated.

{\bf The model.}
To realize our scenario we need two pre-requisites, \cite{OA}.
First, we need two channels,
a FV channel, to drive inflation in the parent Universe, 
and a TV channel, to drive a shorter 
- but well appreciable - inflation inside the bubbles.
Second, we need a tunneling rate  tunable in time, in order to
produce a nucleation peak at the right epoch. Our model 
has just these features: it is certainly not the unique possibility,
but it is a rather simple and geometrically intuitive one.

The model works in fourth order gravity \cite{STARO}, and
 exploits two fields: one, Starobinsky's scalaron $R$ 
 (i.e. the Ricci scalar)
drives the slow-rolling inflation; the second, $\psi$, performs
the first-order phase transition. The phase transition
dynamics is governed both
by the potential of $\psi$ and by its coupling to $R$; the dynamics
of the slow-roll is ``built-in'' in the fourth order Lagrangian:
 ${\cal L}_{grav}=-R+R^2 /  6 M^{2}W(\psi)\;$ 
($c=\hbar=G=1$); the matter Lagrangian is instead standard
and contains the usual potential $  V(\psi) $.
The coupling of the scalaron with $\psi$ can be thought of as
  a field-dependent effective mass
	$M_{eff}(\psi)=M W^{1/2}(\psi)\approx M \,, $
just like the Brans-Dicke scalar-tensor coupling is a field-dependent
Planck mass, and remains hidden in the Yukawa corrections \cite{STELLE}
to Newton's gravity at $10^5 \div 10^6$ Planck lengths 
\cite {STARO} \cite{ALO}. 
We are interested only in the last $N_T \,>\, N_H$ $e$-foldings  
of inflation.
This theory  can be conformally transformed \cite{W} into
canonical gravity:
$\tilde g_{\alpha\beta} =e^{2\omega}g_{\alpha\beta}\,,
e^{2\omega}=|\partial {\cal L} / \partial R|
=1-  R / 3M_{eff}^{2} \;. $ Then, it becomes
 undistinguishable from  Einstein gravity  with two fields 
 $\psi$ and $\omega$, coupled by a potential  
	$U(\psi,\omega)$ linear both in $V(\psi)$ and in $W(\psi)$.

The {\it ansatz} of a quartic  $ W(\psi)=1+8\lambda \psi^2
       (\psi-\psi_0)^2/\psi_0^4 \,, $
  with two degenerate vacua at 0 and $\psi_0$
and a mass term   $ V(\psi)=m^{2}\psi^{2}/2\, $
realizes the two conditions discussed above:	
it carves in fact   two parallel channels of different 
height, separated by a peak of height $\lambda/2$ at $\psi_{PK}=\psi_0/2$.
The degeneracy of $W(\psi)$ in $\psi=0$ and $\psi=\psi_0$
is  removed by $V(\psi)$;
the TV channel
remains exactly at $\psi_{TV}=0$, while the FV channel 
remains approximately  at $\psi_{FV}=\psi_0$.

Let us now follow the evolution  of a generic bubble nucleated at $N$.
After nucleation, the bubble slow rolls down the TV channel for
$N$ $e$-foldings, and then exits inflation, reaching the global
minimum at $\omega=\psi=0$, where it reheats and enters its
Friedmannian radiation dominated era (RDE). At the same time, the
external space also slow rolls for  nearly $N$ $e$-foldings down the FV
channel. After this first inflation, however, the external space 
undergoes a second inflation along the $\psi$-direction, and for
$\omega\approx 0$. This second inflation lasts approximately 
$N_2=2\pi \psi_0^2$ $e$-foldings. Only then does the external space 
reach its RDE. This second inflation
is crucial in our model. In fact, we know that the
bubble's walls grow, as seen from inside, at the velocity of light
as long as the external space
is deSitter. During the first inflation, when both the bubble and
the external space are deSitter,  the walls expand comovingly;
the bubble comoving size is then $L\approx L_H e^{N-N_H}$. During the second
inflation, when the bubble is in  RDE, the  causal horizon expands
overcomovingly as $a^2$, where $a$ is measured from the inside. 
At the end of this
era the bubble  size acquires therefore an extra factor of $a$
becoming $L\approx L_H e^{N-N_H} \times e^{N_2}$.
When finally both the inside and the outside of the bubble are 
in the Friedmannian
regime, the bubble's walls expand again comovingly (as long as  they are of
superhorizon size): now, however, the causal horizon  is faster than 
 comoving expansion, and the walls may become visible. The relevance of the
second inflationary stage is that it allows the bubble comoving size  
to become  as large as
wanted by tuning $N_2$; in particular, the bubble can be  yet invisible
because many times larger than our present horizon. Unlike the other models,
in which we will never see the walls, because the external space is always 
deSitter,
in our model we will see the walls in the future. How remote is this
future depends on $N_2$. Since we want $N \approx N_H$,
it is enough to choose $N_2$  larger than, say, 100, to ensure 
that the bubble walls  are well outside our present horizon. 
This implies a very reasonable $\psi_0>4$.

 With these general considerations in mind, we can proceed
now to work out the details of  the nucleation process, \cite{COLEMAN}.
The tunneling rate $\Gamma$ can be written  as 
	$\Gamma={\cal M}^4 \exp\,(-S)\,, $
where ${\cal M}$ is of the order of the energy of the false vacuum, and $S$
is the minimal Euclidean action, i.e. the action for the 
so-called bounce solution of the Euclidean equation of motion.
For the calculation of $S$
 we can use directly standard physics \cite{COLEMAN}
provided we satisfy the  thin wall limit (TWL), which is not difficult
to achieve.   The result is  \cite{OA} 
\begin{equation} \label{SEN1}
S=\left(N / N_1\right)^4 \,,\quad
 N_1 \approx \sqrt{m^3\psi_0 / M^2\lambda}\,.
\end{equation}
Thus $N\gg N_1$ to avoid spinodal decomposition \cite{TWW}. 
In particular  we will exploit the fact that   $S$ decreases as $N^4$
(bubble nucleation  more likely later  than  earlier)
and increases as  $\lambda^2$ (a quenching opportunity).
Finally,  the relevant parameter   \cite{KOLB}
$     Q=4\pi \Gamma / 9H^4 \,,  $ 
which measures  the 
number of bubbles  per horizon volume per Hubble time 
 can be written  as \cite{OA}
	\begin{eqnarray}\label{BASIC} 
	Q(N)&=&	\exp  \left( (N_0^4-N^4 ) / N_1^4 \right )\,, \nonumber\\
	\left( {\cal M} /  M_{eff} \right)^4&=& 
(9 / 64 \pi) \exp \left(N_0 / N_1\right)^4 \, .
	\end{eqnarray} 
Thus $N_0$   tells us when the physics is being done because
 it estimates the peak of the bubble spectrum.

To summarize, our model has four free parameters: $M$,
setting the slow-rolling inflationary rate;
$m$, setting the energy difference  between vacua;
$\lambda$, setting the barrier height; and finally $\psi_0$,
setting the separation between vacuum channels. These constants 
completely define 
the slow-rolling and the phase transition dynamics:
we should fix all four of them, but 
for the time being we fix only the two combinations thereof we have called
$N_0$ and $N_1$; we hope the remaining freedom will suffice to meet
forecoming constraints.
Furthermore, there is one feature we have to insert by hand, 
it is the mechanism
by which the bubble production is halted {\it ex abrupto}.
 We have chosen to do this by
inserting {\it ad hoc}
at the desired point, $N=N_E$,  a sudden ramp in $\lambda$ 
which then becomes effectively $\lambda(\omega)$. The detailed form
of  $\lambda(\omega)$ is not important as long as 
the  increase in  $S$ is sharp and sudden enough to quench
instantaneously the bubble production.  In fact, given $N_E$ as said above,
we find  $N_0$  by fixing 
the fraction $X$ of the FV phase that we want to turn into the TV phase
through (\ref{X}) below.	
The visual difference with  \cite{OA} consists
in the fact that the FV channel here does not merge into the TV channel
at some $N>0$, but plunges onto it perpendicularly only  at $N\approx 0$. 
Therefore, as already mentioned, whatever remains in the FV phase
slow-rolls classically over  a double inflationary path \cite{DI},
an  essential feature in our model.

We now proceed to the evaluation of the tunneling probability \cite{GW}:
	$dn / dt=\Gamma V_{FV}\,, $ 
 where $dn$ is the number of bubbles per horizon 
nucleated in the time interval $dt$,  
 $V_{FV}=a^3 (4\pi / 3 H^3) \exp (-I)\,,$  and
 $ I(t)=(4\pi / 3 ) 
\int_{-\infty}^{t} dt'\Gamma(t')
\left(a(t')\int_{t'}^{t} dt'' / a(t'')  \right)^3 \,.$

 Incidentally the fact that  $dn/dt$ is proportional 
 to the FV volume left at the time $t$ (i.e. the volume not already 
 occupied by bubbles) implies that a turnaround
 is possible 
 in the bubble production. This is because, due to the  doubly exponential
  nature of the process, after a certain time the FV volume fraction 
  may decrease faster than the product $a^3\Gamma$ increases.
 This turnaround would  indicate that the transition
 is being completed:  again, in this paper, unlike \cite{OA},
 we interrupt  the transition  sharply  at $N_E$
 just before all this happens, \cite{LIN}.

Now,  bubble spectra can be obtained either through numerical integrations   
(see Fig. 2 in \cite{OA} where for the first time realistic bubble
spectra were given) or, better, 
 through a working analytic approximation, which is necessary to
 understand the complex role of our four parameters. Algebraic details
 will be given in future work \cite{ABO}, alongside with further applications.
 Here we outline the procedure: firstly we change from one to the other of the
four equivalent time variables at our disposal, 
$t\,, N\,,L\,,$ and $\Omega_0$ where $L$ is the scale associated with $N$,
as seen from the inflating background, $ L=L_H \exp\, (N-N_H)\,,$
and $\Omega_0$ is given in (\ref{OMEGA0}), by writing
$  dn / dL = \, (dn / dN )/L\, = \, -(dn  / dt ) /HL \,. $
Secondly we get the fraction of space in bubbles of size $L$,
$(dP / dL) = \left(L / L_T  \right)^3 (dn  / dL ) \,,  $
and thirdly the fraction of space in bubbles of a given $\Omega_0$,
$  (dP / d\Omega_0)=(dP / dN )\left(d\Omega_0 / dN \right)^{-1}   \,.  $
Fourthly we evaluate the fraction of space in useful bubbles,
 \begin{equation} \label {X}
X(N_0,N_E)=\int^{N_E}_{\infty}(dP / dN) dN  \;.
\end{equation}

In Fig. 1 we give five examples of spectra obtained with $N_1 = 21$
from four values of $N_0$ decreasing from top to bottom, as shown.
The leftmost dotted curve, $N_0=61$, achieves the completion of
the phase transition, $N_E=0$, and is hence labelled with $X=1$.
This curve is important  because it achieves turnaround 
approximately at $\Omega_0)=.2$ or $N=60$: hence, when the break
in the bubble production is introduced at $N_E=60$, the same curve,
now shown as the uppermost solid line with a vertical cut, yields
a truncated spectrum with $X=.76$ and the peak where needed.
{\it A fortiori}, lower values of $N_0$, which would achieve 
later turnarounds, attain lower values of $X$ as shown, provided
 the break is kept in place, but continue to peak at $\Omega_0=.2$.

{\bf Conclusions.}
We contributed one special toy model to the lore of the flat, inflationary 
Universe filled to a non-negligible fraction by super-horizon-sized 
underdense bubbles, which approximate open Universes. 
This of course reconciles the astronomical observations
in favour of $\Omega_0 \approx 0.2$ with inflation. 
Our own bubble-Universe is one
of an infinite number of similar bubbles. 
Contrary to the single-bubble scenario \cite{GW}
\cite{JAP} \cite{BGT} and 
the many-bubble model of Ref. \cite{LIN}, in our model the external space
also ends inflation a tunable number of $e$-foldings
after the bubbles enter their RDEs; 
the bubbles themselves reenter the horizon in the distant future.
The interesting features of our model are that
 i) we can tune the parameters to achieve maximal probability 
for the nucleation of TWL bubbles around any observed $\Omega_0<1$  
without assuming special
initial conditions and without destroying the CMB isotropy, and that
ii) this probability is not infinitesimally small.

It is worth remarking again that the
measure of $\Omega_0$ along with the assumption that the Universe had an
inflationary epoch, and that our position is  generic, puts
strong constraints on the shape and on the fundamental parameters of
the primordial potential and eventually will fix them,
although for the time being we are limited to such combinations thereof like
$N_0$ and $N_1$.

Inside each bubble one has the usual mechanism of generation of
inflationary perturbations \cite{RP} \cite{JAP} \cite{BGT}. 
It is then possible that reducing the
local $\Omega_0$ to $0.2$ is sufficient to reconcile canonical CDM with
large scale structure. 
 However, evidences are increasing toward the
presence of huge voids in the distribution of matter in the
present Universe, and for velocity fields that are difficult to
explain without a new source of strong inhomogeneities. 
If this were the case, the need may arise of an additional 
primordial phase transition occurring  50 or so $e$-foldings 
before the end of inflation, exactly like in Ref. \cite{OA}.

{\bf Acknowledgments}. We are indebted to A. Linde and A. Mezhlumian
for correspondence after the preprint version of this paper
and to countless friends for patient listening and sharp speaking.


{\bf Fig. 1}  Peaks in the bubble spectra of a bubbly Universe.


\begin{references}
\bibitem{BLD}
N.A. Bahcall, L.M. Lubin, \& V. Dorman,	
  Astrophys. J. {\bf  447}, L81 (1995). 
	\bibitem{CFA2}
	C. Park, M.S. Vogeley, M. Geller, \& J. Huchra,    
  Astrophys. J. {\bf  431}, 569 (1994).
  \bibitem{DEFW}
M. Davis, G. Efstatiou, C. Frenk, \& S.D.M. White,
 Nature {\bf 356} 489 (1992).
  \bibitem{PEE}
  P.J.E. Peebles, Principles of Physical Cosmology,
  (Princeton Univ. Press, 1993).
  	\bibitem{BHM}
  M. Bolte \& C.J. Hogan, Nature {\bf 376} 399 (1995);
  J. Maddox, Nature {\bf 377} 99 (1995).
	\bibitem{RP}
	B. Ratra  \&  P.J.E. Peebles  , Astrophys. J. {\bf 432}
	L5 (1994); preprint PUPT, A-1444
  	\bibitem{OS}
  J.P. Ostriker \& P.J. Steinhardt, Nature {\bf 377} 600 (1995).
	\bibitem{Gaeta}
A.A. Starobinsky, paper presented at {\it Very Early Universe}, Gaeta, 1995.
	\bibitem{MST}
M.S. Turner, G. Steigman \& L.L. Krauss, Phys. Rev. Lett. {\bf 52} 2090 
(1984). 
	\bibitem {Pogosyan}
D.Yu. Pogosyan \&  A.A. Starobinsky in {\it Birth of the Universe \& 
Fundamental
Physics} p.195 (Springer: Heidelberg) F.Occhionero ed.;
M. Davis, F.J. Summers \& D. Schlegel,	Nature {\bf 359} 393 (1992).
	\bibitem{KTF}
	A. Kashlinsky , I. Tkachev , \& J. Frieman 
	 Phys. Rev. Lett. {\bf 73} 1582 (1994).
	\bibitem{COLEMAN}
S.Coleman,  Phys. Rev. D  {\bf 15} 2929 (1977); C. Callan and S. Coleman,
{\it  ibid.} {\bf 16} 1762 (1977);
 S. Coleman and F. De Luccia, {\it ibid.} {\bf 21} 3305 (1980).
	 \bibitem{GOTT}	
J.R. Gott III, Nature {\bf 295} 304 (1982);
J. R. Gott III \& T.S. Statler, Phys. Lett. {\bf 136B} 157 (1984). 	
	\bibitem{GW}
	A. Guth \& E. Weinberg B.,  Nucl. Phys.,  
	 {\bf B212} 321 (1983)
	\bibitem{JAP}
	M. Sasaki, T. Tanaka , Yamamoto K. \& J. Yokoyama,
	Phys. Lett. ,{\bf B317} 510 (1993)
	K. Yamamoto, M. Sasaki, \& T. Tanaka, preprint KUNS 1309
	\bibitem{BGT}
	M. Bucher, A.S. Goldhaber, \& N. Turok,
      Princeton University preprint, hep-th/9411206
	 \bibitem{LIN}
	 A. Linde, Phys. Lett. {\bf B351} 99 (1995);
	 A. Linde, preprint SU-ITP-95-15, gr-qc/9508019;
	 A. Linde \& A. Mezhlumian, preprint SU-ITP-95-11 and 
	 private communication (1995).
	\bibitem{AF}
F.C. Adams \& K. Freese  Phys. Rev. D {\bf 43}  353 (1991).
	\bibitem{DI}
L. Kofman, A. Linde \& A. Starobinsky,   Phys. Lett. {\bf B157} 361 (1985);
L. Amendola, F. Occhionero \& D. Saez, Astrophys. J. {\bf 349} 399, (1990);
S. Gottl\"ober, V. M\"uller \& A. Starobinsky,  Phys. Rev. 
D {\bf 43}  2510 (1991).
	\bibitem{OA}
F. Occhionero \& L. Amendola,  Phys. Rev. D {\bf 50} 4846 (1994).
	 \bibitem{AO}
L. Amendola \& F. Occhionero, Astrophys. J. {\bf 413} 39 (1993).
	\bibitem{TWW}
M.S. Turner, E.J. Weinberg \& L. M. Widrow, Phys. Rev. D {\bf 46} 2384 
(1992).
	  \bibitem{KT}
E.W. Kolb \& M. Turner, The Early Universe (Addison- Wesley, 1990)
	\bibitem{STARO}
	A.A. Starobinsky,  Sov. Phys. JETP Letters {\bf 30}, 682 (1979).
	\bibitem{STELLE}
K. Stelle, Gen. Rel. Grav. {\bf 9} 353 (1978).
	\bibitem{ALO}
L. Amendola, M. Litterio \& F. Occhionero, Phys. Lett. {\bf 231B} 43 (1989).
	\bibitem{W}
	B. Whitt, Phys. Lett. {\bf 145B}, 176 (1984).
	\bibitem{KOLB}
	E.W. Kolb, Physica Scripta {\bf T36} 199 (1991).	
	\bibitem{ABO}
L. Amendola, C. Baccigalupi \& F.Occhionero, in preparation (1996).

\end{references}
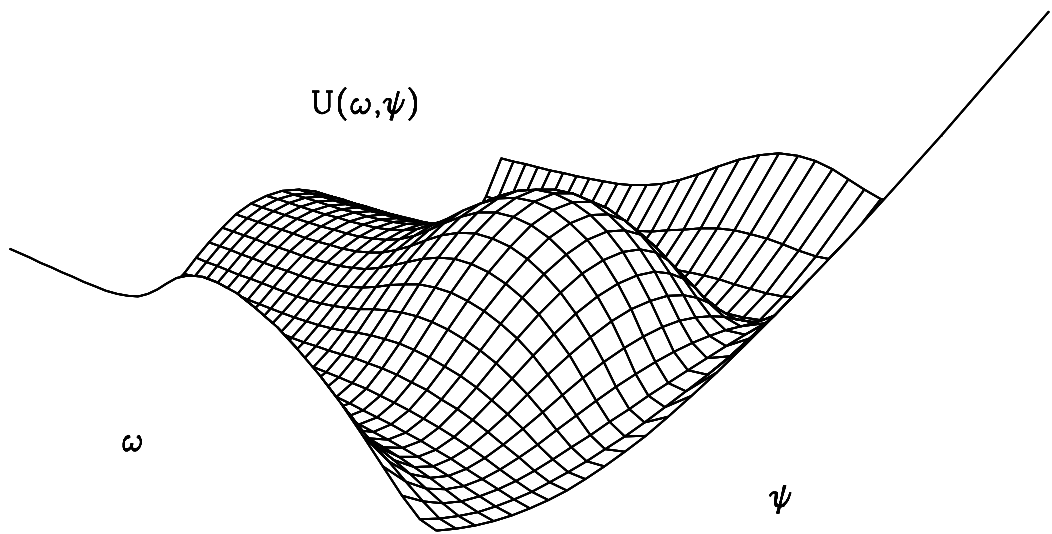
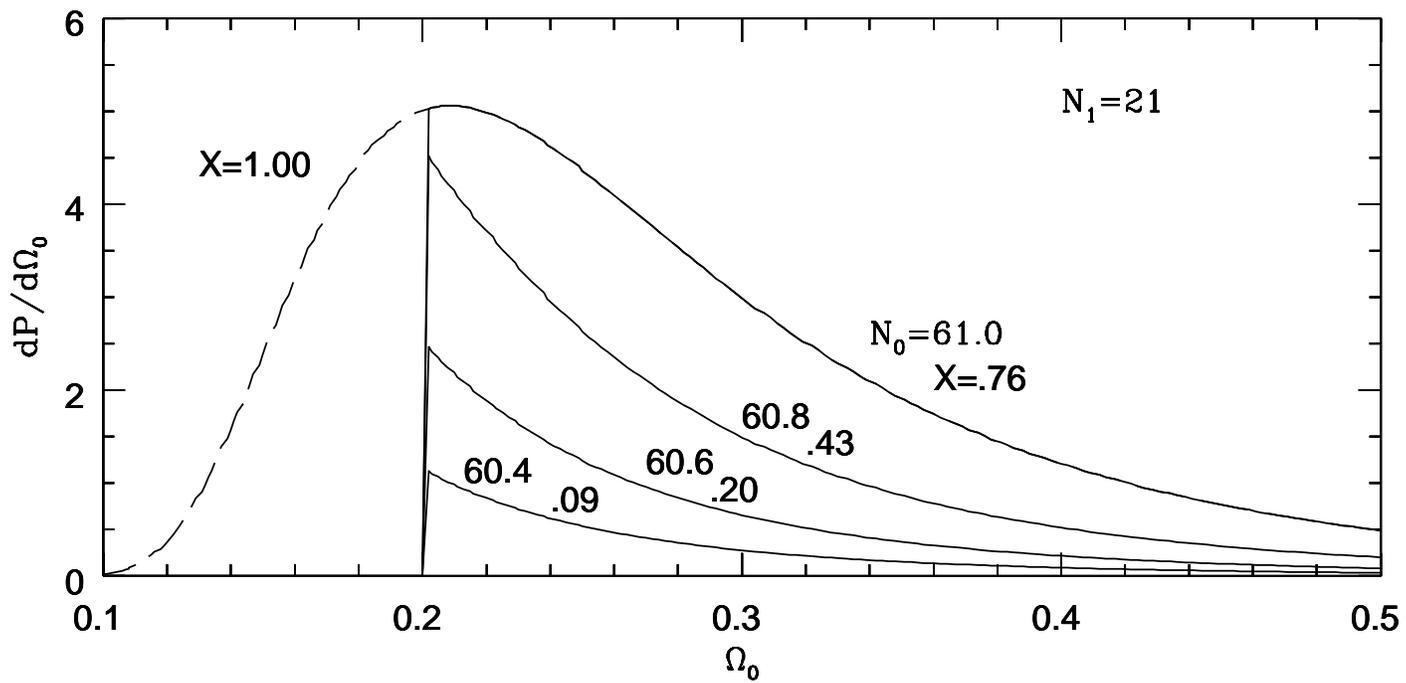
\end{document}